\documentclass[twocolumn,showpacs,preprintnumbers,amsmath,amssymb]{revtex4}
\usepackage{graphicx}% Include figure files
\usepackage{dcolumn}% Align table columns on decimal point
\usepackage{bm}% bold math

\begin{document}

%\preprint{APS/123-QED}

\title{The ground state entanglement in the $XXZ$ model}

\author{Shi-Jian Gu$^{1}$}
\author{Guang-Shan Tian$^{1,2}$}
\author{Hai-Qing Lin$^{1}$}
\affiliation{$^1$Department of Physics, The Chinese University of
Hong Kong, Hong Kong, China} \affiliation{$^2$School of Physics,
Peking University, Beijing 100871, China}

\begin{abstract}
In this paper, we investigate spin entanglement in the $XXZ$ model
defined on a $d$-dimensional bipartite lattice. The concurrence, a
measure of the entanglement between two spins, is analyzed. We
prove rigorously that the ground state concurrence reaches maximum
at the isotropic point. For dimensionality $d \ge 2$, the
concurrence develops a cusp at the isotropic point and we
attribute it to the existence of magnetic long-range order.
\end{abstract}

% 03.67.Mn Entanglement production, characterization,
%          and manipulation (see also 03.65.Ud Entanglement and quantum nonlocality;
%          for entanglement in Bose-Einstein condensates, see 03.75.Gg)
% 03.65.Ud Entanglement and quantum nonlocality
%         (e.g. EPR paradox, Bell's inequalities, GHZ states, etc.)
%         (for entanglement production in quantum information, see 03.67.Mn;
%          for entanglement in Bose-Einstein condensates, see 03.75.Gg)
% 05.70.Jk Critical point phenomena
% 75.10.Jm Quantized spin models

\maketitle

Entanglement, as the exhibition of pure quantum correlations
between separate systems, has become one of the trademarks of the
quantum mechanics for its nonlocal
connotations\cite{ABinstein35,MANielsenb}. Recently, many
physicists have made great efforts to understand the quantum
entanglement in the ground states of some many-body spin models
\cite{AOsterloh2002,TJOsbornee,IBose02,GVidal2003,SJGu03,JVidal04,LAWu04,MFYang04,YChen04}.
One expects that a thorough investigation on the entanglement in
these systems will provide new insight into the quantum phase
transition in these systems \cite{Sachdev}. For example, Osterloh
{\it et al}\cite{AOsterloh2002} studied the concurrence, a measure
of entanglement of two qubits\cite{Hill}, between two spins
located on a pair of nearest-neighbor sites in the
transverse-field Ising model\cite{TJOsbornee}. They found that
this quantity shows singularity and obeys the scaling law in the
vicinity of the quantum phase transition point of the system. On
the other hand, for other models, such as the antiferromagnetic
$XXZ$ chain, the concurrence behaves in a completely different
way\cite{SJGu03}. As shown by Ref. \cite{SJGu03}, the concurrence
is a continuous function of the anisotropic parameter and reaches
its maximum at the transition point. Therefore, in both cases, one
observes that the concurrence itself manifests interesting
behaviors at the quantum phase transition points. However, we
should emphasize that, such behaviors alone may not always signal
a quantum phase transition, as pointed out by the authors of
Refs.~\cite{LAWu04,MFYang04}.

In Ref.~\cite{SJGu03}, we studied extensively nearest-neighbor
spin entanglement in the antiferromagnetic $XXZ$ chain. By
applying results derived from the Bethe ansatz solution of the
model, we showed clearly that the concurrence between two spins
located on a pair of nearest-neighbor sites in the system is a
continuous function of the anisotropic coupling parameter and
becomes maximal at the isotropic Heisenberg point. In this paper,
we continue our discussions on this issue. Our main purpose is to
show that some fundamental properties of the $XXZ$ model, such as
non-degeneracy and concavity of the ground state energy of the
system at the phase transition point, commands strongly on the
behavior of the concurrence. Therefore, we expect that the same
scenario will appear in a wide class of localized spin models,
such as the spin ladder model and, in particular, the $XXZ$ model
in higher dimensions~\cite{OFSyljuasenm}. It is well known that,
as far as the above-mentioned properties are concerned, the ground
states of these models are akin to the antiferromagnetic $XXZ$
chain.

This paper contains two parts. In the first part, based on some
well-known facts about the antiferromagnetic spin models, we prove
rigorously that, when the antiferromagnetic $XXZ$ model is defined
on a $d$-dimensional {\it finite} bipartite lattice, the
concurrence between two spins located on a pair of
nearest-neighbor sites is an analytical function of the
anisotropic parameter and takes on its maximum at the Heisenberg
isotropic point. Then, in the second part of this paper, justified
by the existence of magnetic long-range order (LRO) in the $XXZ$
model, we use the spin-wave theory to show that a cusp-like
behavior of the concurrence develops in the thermodynamic limit
when the dimensionality of the lattice $d\ge 2$.

To begin with, we first introduce several notations. On a finite
$d$-dimensional simple cubic lattice $\Lambda$ with
$N_\Lambda=L^d$ sites, the Hamiltonian of the antiferromagnetic
$XXZ$ model is
\begin{equation}
\hat{H}_{XXZ} = \sum_{\langle{\bf ij}\rangle}
\left(\hat{S}_{\bf i}^x \hat{S}_{\bf j}^x
+ \hat{S}_{\bf i}^y S_{\bf j}^y + \Delta \hat{S}_{\bf i}^z
\hat{S}_{\bf j}^z\right),
\end{equation}
where $\hat{S}_{\bf i}^x,\>\hat{S}_{\bf i}^y$ and $\hat{S}_{\bf
i}^z$ are spin-1/2 operators at site $\bf i$ and
$\Delta=J_z/J_{x}\; (J_x=J_y)$ is a dimensionless parameter
characterizing the anisotropy of the model. The sum in the
Hamiltonian is over all pairs of nearest-neighbor sites $\bf i$
and $\bf j$. Obviously, this Hamiltonian commutes with the total
spin $z$-component operator $\hat{S}^z_{\rm total}=\sum_{\bf
i}\hat{S}^z_{\bf i}$. Thus, each eigenstate of the Hamiltonian is
also an eigenstate of $\hat{S}^z_{\rm total}$. Consequently, the
Hilbert space of the system can be decomposed into numerous
subspaces $V(M)$. In each subspace, the spin number
$\hat{S}^z_{\rm total}=M$ is specified. It is well known that, on
a finite simple cubic lattice $\Lambda$, the ground state of the
$XXZ$ model is nondegenerate in any admissible subspace $V(M)$
\cite{Lieb,Affleck}. In particular, its global ground state
$\Psi_0(\Lambda,\>\Delta)$, which coincides with the ground state
of the model in the subspace $V(M=0)$ \cite{Affleck}, is also
nondegenerate. Therefore, all the physical quantities, such as the
ground state energy $E_0(\Lambda,\>\Delta)$ and the spin
correlation function $\langle\hat{S}_{\bf i}^z\hat{S}_{\bf
i}^z\rangle$ are analytical functions of the parameter $\Delta$,
as long as the lattice is finite.

  The conservation of $\hat{S}_{\rm total}^z$ implies also that,
with respect to the standard basis vectors
$\vert\uparrow\uparrow\rangle$,
$\vert\uparrow\downarrow\rangle$,
$\vert\downarrow\uparrow\rangle$ and
$\vert\downarrow\downarrow\rangle$,
the reduced density matrix of two spins on a pair of nearest-neighbor
lattice sites $\bf i$ and $\bf j$ can be put
into the following block-diagonal form
\begin{equation}
\hat{\rho}_{\bf ij} =
\left(
\begin{array}{llll}
u^+ & 0 & 0 & 0 \\
0 & w_1 & z & 0 \\
0 & z^* & w_2 & 0 \\
0 & 0 & 0 & u^-
\end{array}
\right).
\end{equation}
As a result, the concurrence of the two spins is $C_{\bf
ij}=2\max\left(\vert z\vert-\sqrt{u^+
u^-},\>0\right)$\cite{KMOConnor2001}. In terms of the correlation
function $G_{\bf
ij}^{\alpha\alpha}=\langle\hat{S}^\alpha\hat{S}^\alpha\rangle,\>
\alpha=x,y,z$, it can be explicitly written as\cite{XWang2002PLA}
\begin{eqnarray}
C_{\bf ij} =2\max\left(\left| G_{\bf ij}^{xx} + G_{\bf ij}^{yy}\right|
- G_{\bf ij}^{zz} - \frac{1}{4},\>0 \right).
\label{eq:concurrence}
\end{eqnarray}
By the variational principle, one can show that all the spin
correlations functions $G_{\bf ij}^{\alpha\alpha}$ are negative.
Thus, one has
\begin{eqnarray}
C_{\bf ij}
& = &
\left(- G_{\bf ij}^{xx} - G_{\bf ij}^{yy} - \frac{1}{4}
- G_{\bf ij}^{zz}\right) \nonumber\\
& = &
\left(-\epsilon_{\bf ij}^0(\Lambda,\>\Delta)
- \frac{1}{4} + (\Delta - 1) G_{\bf ij}^{zz}\right),
\end{eqnarray}
where $\epsilon_{\bf ij}^0(\Lambda,\>\Delta)=
E_0(\Lambda,\>\Delta)/N_B$ ($N_B$ is the number of bonds in the
lattice) is the ground state energy density per bond. Furthermore,
since all quantities in $C_{\bf ij}$ are analytical functions of
the parameter $\Delta$, we are allowed to take derivatives of it
with respect to $\Delta$. In particular, after taking the first
order derivative of $C_{\bf ij}$, we obtain
\begin{equation}
\frac{\partial C_{\bf ij}}{\partial\Delta}
= 2\left(- \frac{\partial\epsilon_{\bf ij}^0(\Lambda,\>\Delta)}
{\partial\Delta} + G^{zz}_{\bf ij}
+ \left(\Delta-1\right) \frac{\partial G^{zz}_{\bf ij}}
{\partial\Delta} \right).
\label{Concurrence Derivative}
\end{equation}
Again, due to the nondegeneracy of the global ground state
$\Psi_0(\Lambda,\>\Delta)$ of the $XXZ$ model on a finite lattice,
we can use the Hellman-Feynman theorem to calculate the derivative
$\partial\epsilon_{\bf ij}^0(\Lambda,\>\Delta)/\partial\Delta$,
which equals to $G^{zz}_{\bf ij}$. Therefore, we finally obtain
\begin{eqnarray}
\frac{\partial C_{\bf ij}}{\partial\Delta}
& = &
2 \left(\Delta-1\right) \frac{\partial G^{zz}_{\bf ij}}
{\partial\Delta} \nonumber\\
& = &
2 \left(\Delta-1\right) \frac{\partial^2
\epsilon_{\bf ij}^0(\Lambda,\>\Delta)}
{\partial\Delta^2}.
\label{First Derivative}
\end{eqnarray}
Immediately, one sees that $\Delta=1$ is an extreme point of the
concurrence.

Next, we show that $\Delta=1$ is actually a maximal point of
$C_{\bf ij}$ and the concurrence does not have other extreme
point. In fact, both the statements are the corollaries of
concavity of the ground state energy $E_0(\Lambda,\>\Delta)$ of
the Hamiltonian $\hat{H}_{XXZ}$ with respect to the anisotropic
parameter $\Delta$. By the variational principle\cite{Franklin},
we know that, for any two parameters $\Delta_1$ and $\Delta_2$,
the following inequality
\begin{equation}
E_0(\Lambda,\>\lambda\Delta_1 + (1 - \lambda)\Delta_2)
\ge \lambda E_0(\Lambda,\>\Delta_1)
+ (1 - \lambda) E_0(\Lambda,\>\Delta_2),
\label{Concavity}
\end{equation}
where $0\le\lambda\le 1$, holds true for the ground state energy
$E_0(\Lambda,\>\Delta)$. In particular, when
$E_0(\Lambda,\>\Delta)$ is differentiable with respect to
$\Delta$, inequality (\ref{Concavity}) is equivalent to
\begin{equation}
\frac{\partial^2 E_0(\Lambda,\>\Delta)}{\partial\Delta^2}
\le 0.
\label{Second Derivative}
\end{equation}
Consequently, we have also
$\partial^2\epsilon_{\bf ij}^0(\Lambda,\>\Delta)/\partial\Delta^2\le 0$.
Now, let us take derivative of Eq.~(\ref{First Derivative})
again with respect to $\Delta$. It yields
\begin{equation}
\left. \left.
\frac{\partial^2 C_{\bf ij}}{\partial\Delta^2}
\right|_{\Delta=1}
= 2\frac{\partial^2 \epsilon_{\bf ij}^0(\Lambda,\>\Delta)}
{\partial\Delta^2}\right|_{\Delta=1} \le 0.
\end{equation}
Therefore, $\Delta=1$ is indeed a maximal point of the concurrence.

Finally, we prove that $\Delta=1$ is the unique extreme point of
the concurrence $C_{\bf ij}$. For that purpose, we notice that
inequality (\ref{Second Derivative}) is actually strict. In other
words, the equal sign in it can be ignored. This can be easily
understood by observing the following fact: As $\Delta$ increases
from $-\infty$ to $\infty$, quantity $\langle\hat{S}_{\bf
i}^z\hat{S}_{\bf i}^z\rangle =\partial\epsilon_{\bf
ij}^0(\Lambda,\>\Delta)/\partial\Delta$ becomes more and more
negative. Consequently, the product on the right-hand side of
Eq.~(\ref{First Derivative}) cannot be zero at any point except
$\Delta=1$. That completes our discussion on the general behavior
of the concurrence $C_{\bf ij}$ for the antiferromagnetic $XXZ$
model on a finite $d$-dimensional simple cubic lattice. In
addition, we point out that the above proof can be easily extended
to other cases, such as the spin ladder model at $J_\perp=0$.

In the following, we shall discuss the behavior of the concurrence
in the thermodynamic limit. In Ref.~\cite{SJGu03}, by using the
Bethe ansatz solution of the one-dimensional $XXZ$ chain, we
obtained  the explicit expression of the concurrence near the
isotropic point
\begin{equation}
C_{i,i+1} = C_0 - C_1(\Delta-1)^2, \label{1DXXZ}
\end{equation}
where $C_0$ and $C_1$ are two real constants. Therefore, the
concurrence of the 1D $XXZ$ chain is a differentiable function of
$\Delta$ in the thermodynamic limit. However, things are quite
different in higher dimensions. For the $XXZ$ model in higher
dimensions, there exists no exact solution. One either uses
approximate analytical approach such as the spin-wave theory or
numerical approach such as exact diagonalization studies of finite
lattice. To obtain results in the thermodynamic limit, finite size
scaling analysis must be performed. By using the stochastic series
expansion quantum Monte Carlo method for lattices up to $16 \times
16$, Sandvik~\cite{Sandvik} did an extensive study on the
two-dimensional S=1/2 antiferromagnetic Heisenberg model. The
finite size results for various ground state quantities were
extrapolated to the thermodynamic limit using fits to polynomials
in $1/L$, constrained by scaling forms previously obtained from
renormalization-group calculations for the nonlinear sigma model
and chiral perturbation theory. He demonstrated that the results
were fully consistent with the predicted leading finite size
corrections. With the same scaling forms, Lin, Flynn, and
Betts~\cite{Lin-2DXXZ} studied the $XXZ$ model on square lattices
and obtained various quantities as functions of anisotropic
parameter $\Delta$ for the infinite system. Two conclusions from
pervious works \cite{Sandvik,Lin-2DXXZ,JEHirsch89,WZheng91} are
relevant to the present study: (i) results obtained from the
spin-wave theory are qualitatively correct and quite accurate,
usually within 3 percent as compared with exact solution on finite
lattices; (ii) derivatives of the ground state energy with respect
to the anisotropic parameter $\Delta$ are not continuous at the
Heisenberg point $\Delta=1$, for example, see Figure 3 in Ref.
\cite{Lin-2DXXZ}. This conclusion is consistent with the belief
that there exists antiferromagnetic long-range order (LRO) in the
d-dimensional $XXZ$ model for $d \ge 2$. In fact, the existence of
the LRO for $d \ge 3$ was rigorous proved~\cite{XXZLRO}, while for
$d=2$ most numerical studies support it. Based on these
conclusions, we apply the spin-wave theory to calculate the
concurrence $C_{\bf ij}$ of the $XXZ$ model. We also use exact
diagonalization results as complementary. As shown in the
following, the symmetry breaking in the thermodynamic limit, which
is absent in the one-dimensional case, causes the singular
behavior of the concurrence at the quantum phase transition point.

Following the standard procedure, the $XXZ$ Hamiltonian is mapped
into a boson model via the Holstein-Primakoff (HP) transformation
\begin{eqnarray}
&&\hat{S}^+_{\bf i}
=\sqrt{2S}\,(1-\hat{n}_{\bf i}/2S)^{1/2}\hat{a}_{\bf i}
\simeq \sqrt{2S}\,(1-\hat{n}_{\bf i}/4S)\, \hat{a}_{\bf i},\nonumber \\
&&\hat{S}^-_{\bf i}=\sqrt{2S}\>
\hat{a}^\dagger_{\bf i}(1-\hat{n}_{\bf i}/2S)^{1/2}
\simeq \sqrt{2S}\,\hat{a}^\dagger_{\bf i}(1-\hat{n}_{\bf i}/4S), \nonumber \\
&&\hat{S}^z_{\bf i}=S- \hat{a}_{\bf i}^\dagger \hat{a}_{\bf i},
\label{HP Transformation}
\end{eqnarray}
where $\hat{a}_{\bf i}$ and $\hat{a}_{\bf i}^\dagger$ are boson
creation and annihilation operators at site $\bf i$ for the spin
deviation. In the region $\Delta >1$, the antiferromagnetic
ordering is in the spin-$z$ direction. Consequently, we have
\begin{eqnarray}
\hat{H}_{XXZ}/\Delta
& \simeq &
\sum_{\langle{\bf ij}\rangle} \left[-S^2
+ S \left(\hat{a}_{\bf i}^\dagger\hat{a}_{\bf i} + \hat{a}_{\bf
j}^\dagger\hat{a}_{\bf j}\right) \right. \nonumber\\
& + & \left. x S \left(\hat{a}_{\bf i}\hat{a}_{\bf j} +
\hat{a}_{\bf i}^\dagger\hat{a}_{\bf j}^\dagger\right)\right],
\end{eqnarray}
where $x=1/\Delta$. Using Fourier transform, we rewrite the
Hamiltonian as
\begin{eqnarray}
\hat{H}_{XXZ}/\Delta = -\frac{z}{2}NS^2
+zS\sum_{\bf k} \hat{H}({\bf k}),
\end{eqnarray}
where $z$ is the coordination number of the lattice and
\begin{eqnarray}
\hat{H}({\bf k})= \hat{a}_{\bf k}^\dagger \hat{a}_{\bf k}
+\frac{x\gamma_{\bf k}}{2}\left(\hat{a}_{\bf k}\hat{a}_{-{\bf k}}
+ \hat{a}_{-{\bf k}}^\dagger \hat{a}_{\bf k}^\dagger\right)
\end{eqnarray}
with $\gamma_{\bf k}=\frac{2}{z} \sum_{m=1}^d\cos k_m$. By
applying the Bogoliubov transformation
\begin{eqnarray}
&&\hat{a}_{\bf k}= u_{\bf k} \hat{c}_{\bf k}
- v_{\bf k} \hat{c}_{-{\bf k}}^\dagger \nonumber \\
&& \hat{a}_{\bf k}^\dagger = -v_{\bf k}\hat{c}_{-{\bf k}}
+ u_{\bf k}\hat{c}_{\bf k}^\dagger,
\end{eqnarray}
we diagonalize $\hat{H}({\bf k})$ and obtain
\begin{eqnarray}
\hat{H}({\bf k})
& = &
v_{\bf k}^2 - x \gamma_{\bf k}u_{\bf k} v_{\bf k} \nonumber\\
& + &
\left(u_{\bf k}^2 + v_{\bf k}^2 -2x u_{\bf k}
v_{\bf k}\gamma_{\bf k}\right)
\hat{c}_{\bf k}^\dagger\hat{c}_{\bf k},
\end{eqnarray}
where the $u_{\bf k}$ and $v_{\bf k}$ satisfy
the following constraint conditions
\begin{eqnarray}
& &
u_{\bf k}^2 - v_{\bf k}^2=1, \nonumber \\
& &
\frac{x\gamma_{\bf k}}{2}(u_{\bf k}^2
+ v_{\bf k}^2) - u_{\bf k} v_{\bf k} = 0.
\end{eqnarray}
Finally, the ground state energy of the model in the region of
$\Delta>1$ can be written as
\begin{eqnarray}
E_0(\Delta>1) =-\frac{z}{2} N S^2
+ \frac{zS}{2}\sum_{\bf k} \left(\sqrt{1
- x^2\gamma_{\bf k}^2} - 1\right).
\label{Ground-State Energy1}
\end{eqnarray}

By similar approach, we can also obtain the ground state energy of
the $XXZ$ model in the parameter region of $0<\Delta<1$. In this
case, the system has antiferromagnetc order in the $XY$ plane in
the thermodynamic limit. As a result, the diagonalized Hamiltonian
has the following form
\begin{eqnarray}
\hat{H}({\bf k})
& = &
\left(1 + y\gamma_{\bf k}\right)
v_{\bf k}^2 - x\gamma_{\bf k} u_{\bf k} v_{\bf k} \nonumber \\
& + &
\left[\left(1 + y\gamma_{\bf k}\right)
\left(u_{\bf k}^2 + v_{\bf k}^2\right)
- 2x u_{\bf k} v_{\bf k} \gamma_{\bf k}\right]
\hat{c}_{\bf k}^\dagger \hat{c}_{\bf k},
\end{eqnarray}
where $x=(1+\Delta)/2$ and $y=(1-\Delta)/2$, and the corresponding
ground state energy is
\begin{eqnarray}
E_0
& = &
- \frac{z}{2} N S^2 +
\frac{zS}{2}\sum_{\bf k}
\left(1 + y\gamma_{\bf k}\right) \nonumber \\
& \times &
\left(\sqrt{1-x^2\gamma_{\bf k}^2/(1+y\gamma_{\bf k})^2}
- 1\right).
\label{Ground-State Energy2}
\end{eqnarray}

\begin{figure}
\includegraphics[width=7cm]{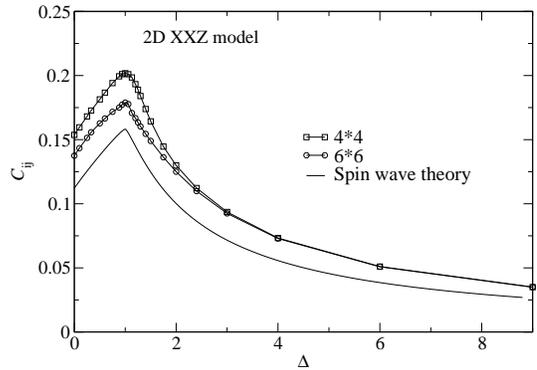}
\caption{\label{figure_xxz2con} The concurrence $C_{\bf ij}$ of
the two-dimensional $XXZ$ model as a function of
$\Delta$($=J_z/J_{x}$). In the Figure, the doted lines are
obtained from the exact diagonalization for $4*4$(square) and
$6*6$(circle) square lattices respectively.}
\end{figure}

\begin{figure}
\includegraphics[width=7cm]{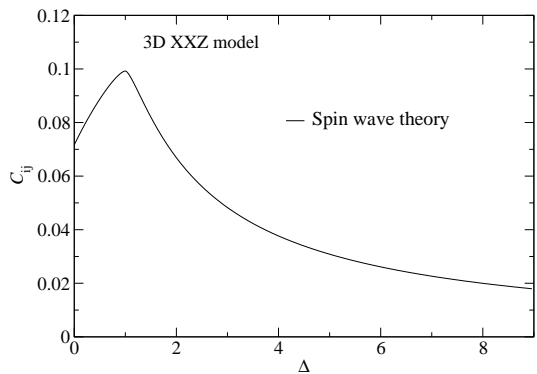}
\caption{\label{figure_xxz3con} The concurrence $C_{\bf ij}$ of
the three-dimensional $XXZ$ model as a function of
$\Delta$($=J_z/J_{x}$).}
\end{figure}

Within the spin-wave theory framework, we calculate the spin
correlation function $G_{\bf ij}^{zz}$ and hence the concurrence
$C_{\bf ij}$ of the model in two and three dimensions. Our results
are shown in Figs.~\ref{figure_xxz2con} and \ref{figure_xxz3con},
respectively. We also show results obtained from the exact
diagonalization of the $XXZ$ model on finite square lattices. The
trend as function of lattice size is clear. It is interesting to
see that, in both cases, the concurrences $C_{\bf ij}$ of the
$XXZ$ model not only have their maximal value at the critical
point $\Delta=1$, but also show discontinuities in their first
derivative with respect to $\Delta$ at the transition point. This
behavior is quite different from the one-dimensional case
(Eq.~\ref{1DXXZ}), as we expected. We attribute this difference to
the existence of the magnetic long-range orders in the system with
$d\ge 2$.

As we have seen, the concurrence $C_{\bf ij}$ is closely related
to the ground state energy of the model. As a result, any
singularity in the ground state energy may be inherited by the
concurrence\cite{LAWu04}. On the other hand, on a finite
$d$-dimensional simple cubic lattice, the ground state of the
antiferromagnetic $XXZ$ model is non-degenerate for
$\Delta\in(-\infty,\>\infty)$\cite{Affleck}. Therefore, the ground
state energy $E_0(\Lambda,\>\Delta)$ as well as the concurrence
$C_{\bf ij}$ are analytical functions of $\Delta$, regardless of
the dimensionality of the lattice. However, it is no longer true
in the thermodynamic limit. For the one-dimensional $XXZ$ model,
it is well known that its ground state in both $\Delta<1$ and
$\Delta>1$ regions does not have magnetic long-range order.
Therefore, we do not expect a dramatic change in the ground state
energy $E_0$ taking place at $\Delta=1$. Consequently, the
concurrence will behave more or less like itself on a finite
lattice. However, in two and three dimensions, the ground state
energy of the system develops a cusp at the transition point in
the thermodynamic limit\cite{Lin-2DXXZ}. This phenomenon can be
understood by the picture of the first excited-energy levels
crossing at $\Delta=1$, required by the existence of magnetic
long-range order~\cite{Tian}. Therefore, a singularity inherited
by the concurrence at the transition point, is expected to appear,
as shown by our calculations.

In summary, we have studied the ground state two-spin
entanglement, as measured by the concurrence, in the
$d$-dimensional $XXZ$ model. We gave a rigorous proof that the
ground state concurrence in the $XXZ$ model reaches maximum at the
isotropic point. We extended our previous studies in one dimension
\cite{SJGu03} to two and three dimensions by using the spin-wave
theory and exact diagonalization technique. The use of the
spin-wave theory is justified by the existence of magnetic
long-range order in the $XXZ$ model for dimensionality $d\ge2$. We
found that the concurrence in two- and three-dimensional $XXZ$
models also reaches maximum at the isotropic point $\Delta=1$.
Unlike the one dimension case, the concurrence shows cusp-like
behavior around the critical point, and its first derivative is
not continuous in the vicinity of the critical point.

This work was supported by a grant from the Research Grants
Council of the HKSAR, China (Project No. 401703). G. S. Tian
acknowledges financial support from the C. N. Yang Visiting
Fellowship program.

Note Added: While we were preparing this paper, we received a
preprint from Dr. M. F. Yang\cite{MFYang04}. Some of our results
were also obtained by him.

\end{document}